\documentstyle[12pt,epsfig]{article}
\newcommand{\be}{\begin{equation}}
\newcommand{\ee}{\end{equation}}
\newcommand{\ba}{\begin{eqnarray}}
\newcommand{\ea}{\end{eqnarray}}

\begin{document}
\renewcommand{\baselinestretch}{1.1} \small\normalsize

\begin{flushright}
{\sc HUB-EP}-98/35\\June 1998
\end{flushright}
\thispagestyle{empty}

\vspace*{1.5cm}

\begin{center}

{\Large \bf Phase structure of U(1) lattice gauge theory\\
with monopole term}

\vspace*{0.8cm}

{\bf Georg Damm$^a$, Werner Kerler$^{b}$}

\vspace*{0.3cm}

{\sl $^a$ Fachbereich Physik, Universit\"at Marburg, D-35032 Marburg,
Germany\\
$^b$ Institut f\"ur Physik, Humboldt-Universit\"at, D-10115 Berlin, Germany}

\end{center}

\vspace*{1.3cm}

\begin{abstract}
\hspace{-2mm}
We investigate four-dimensional compact U(1) lattice gauge theory with 
a monopole term added to the Wilson action. First we consider the phase 
structure at negative $\beta$, revealing some properties of a third phase 
region there, in particular the existence of a number of different states.
Then our present studies concentrate on larger values of the monopole 
coupling $\lambda$ where the confinement-Coulomb phase transition 
turns out to become of second order. 
Performing a finite-size analysis we find that the critical 
exponent $\nu$ is close to, however, different from the gaussian value 
and that in the range considered $\nu$ increases somewhat with $\lambda$.
\end{abstract}

\newpage 

\section{INTRODUCTION}

\hspace{3mm}
The phase transition in 4-dimensional compact U(1) lattice gauge theory
with the Wilson action because of the occurrence of a gap in the energy
histogram is believed to be of first order. Actually this is to be studied
in more detail by a finite-size analysis. Recently critical exponents have
been determined in a higher-statistics study \cite{r97}. There the
critical exponent $\nu$ has been found to decrease towards $\frac{1}{4}$
with increasing lattice size, i.e. towards the value characteristic of
a first-order transition. 

If the Wilson action is extended by a double charge term with coupling 
$\gamma$ the first order transition weakens with decreasing $\gamma$ and
has been conjectured to become of second order at some negative $\gamma$ 
\cite{ejnz85}, which, however, so far has not been confirmed. In 
Ref.~\cite{jln96} instead of the usual periodic boundary 
conditions a spherelike lattice (i.e.~the surface of a 5-cube which is 
homeomorphic to a 4-sphere) has been used. The disappearance of the gap
at $\gamma=0$, -0.2, -0.5 observed there has been interpreted by the authors
as an earlier start of the second order region and the relatively low
value of $\nu$ there as related to a nongaussian fixed point. 

In a more recent investigation of the action with double charge term 
\cite{cct97} the gap has been shown for $\gamma$ down to $-0.4$ to reappear 
on larger spherelike lattices. The absence of a gap on smaller lattices has
been attributed to larger finite-size effects in the spherical geometry. 
These observations are just what was to be expected according to studies
of the influence of different boundary conditions \cite{krw96} where it 
turned out that inhomogeneities weaken the transition and that the gap 
reappears for sufficiently large lattice sizes. 
In addition in Ref.~\cite{cct97} for $\gamma$ between $+0.2$ and $-0.4$
the critical exponent $\nu$ has been found to decrease towards $\frac{1}{4}$ 
with increasing lattice size for toroidal as well as for spherical geometry. 
Further, in some cases stabilization of the latent heat has been observed. 

Thus, for the above actions there are now rather strong indications that, 
at least in the region which has been investigated so far, the transition 
is of first order. Of course, it remains desirable to check this on still 
larger lattices. While this is hardly possible with conventional methods, 
it can be done using the dynamical-parameter algorithm developed in 
\cite{krw94, krw95a}.

After the mentioned evidence of first order the question arises where one 
can find a second-order phase transition in the U(1) gauge system. This 
leads to the modification of the Wilson action 
where a monopole term is added \cite{bs85}, which reads
\be
S=\beta \sum_{\mu>\nu,x} (1-\cos \Theta_{\mu\nu,x})+
\lambda \sum_{\rho,x} |M_{\rho,x}|
\label{a}
\ee
with $M_{\rho,x}=\epsilon_{\rho\sigma\mu\nu}
(\bar{\Theta}_{\mu\nu,x+\sigma}-\bar{\Theta}_{\mu\nu,x}) /4\pi$
where the physical flux $\bar{\Theta}_{\mu\nu,x}\in [-\pi,\pi)$ is 
related to the plaquette angle $\Theta_{\mu\nu,x}\in (-4\pi,4\pi)$ by
$\Theta_{\mu\nu,x}=\bar{\Theta}_{\mu\nu,x}+2\pi n_{\mu\nu,x}$ \cite{dt80}.

Using the action (\ref{a}) from the energy distribution it is seen that the 
gap gets smaller with increasing monople coupling $\lambda$ \cite{krw94,krw95a},
which indicates that the first-order transition gets weaker and 
for sufficiently large $\lambda$ becomes of second order. The latter has been 
corroborated by a finite-size analysis \cite{krw97} which has shown that at 
$\lambda= 0.9$ the critical exponent is already characteristic of a 
second-order transition. 

In addition to showing a second-order phase transition for sufficiently 
large $\lambda$, the action (\ref{a}) is particularly attractive in view of 
the fact \cite{krw94,krw95} that the confinement phase and the Coulomb phase 
are unambiguously characterized by the presence or absence, respectively, of an 
infinite network of monopole current lines, where ``infinite'' on finite 
lattices is to be defined in accordance with the boundary conditions 
\cite{krw96}. Since the probability $P_{\mbox{\scriptsize net}}$ to find
an infinite network takes the values 1 and 0, respectively, it is very 
efficient to discriminate between those phases. For the finite lattices with 
periodic boundary conditions we are considering here, ``infinite'' means 
``topologically nontrivial in all directions''. While for loops this 
characterization is straightforward, to determine the topological properties 
of the occurring networks it is necessary to perform a more elaborate analysis 
based on homotopy preserving mappings \cite{krw94,krw95}. 

In \cite{krw97} the location $\beta_{\mbox{\scriptsize cr}}$ of the
transition from the confinement phase to the Coulomb phase has been determined 
as a function of $\lambda$. It has been found that the related transition line 
continues to negative $\beta$. On the other hand, in Refs.~\cite{hmp94,bhmp96} 
a further transition has been reported at $\beta=-1$ for $\lambda=0$ and at 
$\beta=-0.7$ for $\lambda=\infty$. This suggests to check in more detail
what happens at negative $\beta$.

In the second-order region mentioned above an important question is wether one
has universal critical properties there. The energy distribution 
indicates that this region starts at some finite $\lambda$ above $0.7$ 
\cite{krw94,krw95a}. For $\lambda= 0.9$ the critical exponent $\nu$ is
known to be characteristic of a second-order transition \cite{krw97}. It 
appears important now to get more information on this region.
In particular finite-size analyses at different values of 
$\lambda$ and for a variety of observables are desirable.

In the present work we have performed Monte Carlo simulations to address
the indicated questions. Simulation runs at various values of $\lambda$ and
$\beta$ have been done to get an overview of the situation at negative 
$\beta$. The emerging picture of this is discussed in Sect.~II. The emphasis 
of our investigations has then been on higher-statistics simulations at 
$\lambda=1.1$ and $\lambda=0.8$ in the critical region of the 
confinement-Coulomb phase transition, which have been evaluated by 
finite-size analyses. The results of this are presented in Sect.~III.

\section{PHASE REGIONS}

\hspace{3mm}
In \cite{krw97} the location $\beta_{\mbox{\scriptsize cr}}$ of the
transition from the confinement phase to the Coulomb phase has been determined 
by the maximum of the specific heat $C_{\mbox{\scriptsize max}}$ 
up to $\lambda=1.3$. Because the peak of the specific heat
has been found to decrease strongly with $\lambda$, at larger values 
of $\lambda$ only $P_{\mbox{\scriptsize net}}$ has been used to locate
the transition. In these investigations it has turned out that above 
$\lambda=1.2$ the associated values of $\beta_{\mbox{\scriptsize cr}}$
become negative.

The Wilson action has the symmetry $\beta \rightarrow -\beta$, 
$U_{\Box} \rightarrow -U_{\Box}$ 
\footnote{The transformation $U_{\Box} \rightarrow -U_{\Box}$ of the
plaquette fields can be realized by transforming the link angles as 
$\Theta_{\mu x} \rightarrow \pm \Delta_{\mu x}\Theta_{\mu x}$ where the 
sign choice is according to $\Theta_{\mu x}$ being in the intervals 
$[-\pi,0)$ and $[0,\pi)$, respectively, and where $\Delta_{\mu x}=-1$
if $x_1+x_2+\ldots +x_{\mu -1}$ is odd and $\Delta_{\mu x}=+1$ if it is
even or if $\mu=1$.}.        
At $\lambda=0$ it gives rise to a phase transition at $\beta=-1$ in addition 
to the one at $\beta=1$ \cite{hmp94}. For $\lambda\ne 0$ this
symmetry is violated by the monopole term. At $\lambda=\infty$ only
the transition at negative $\beta$ persists and occurs at about 
$\beta=-0.7$ \cite{hmp94,bhmp96}. 

Here we have checked the occurrence of such transition at negative 
$\beta$ also at intermediate values of $\lambda$ by determining 
$C_{\mbox{\scriptsize max}}$. Our observations indicate a transition 
line extending from $(\lambda,\beta)=(0,-1)$ to $(\infty,-0.7)$. 
In the energy distribution on the $8^4$ lattice we have seen a double-peak
structure at $\lambda=0.5$ and indications of such structures (though
not resolved with present accuracy) at other $\lambda$ values. Together with
the observations at $\lambda=0$ and at $\lambda=\infty$ in 
Refs.~\cite{hmp94,bhmp96} this hints at first order along that transition
line. Of course, it remains to confirm this with higher statistics and on
larger lattices.

We have also investigated some properties of the system below the new 
transition line which will be discussed in the following. Since so far we 
have no indication of a further subdivision of this region, but rather find
similar properties throughout it, we consider it here (at least 
as a working hypothesis) as a third phase.

Figure 1 gives an overview of the phase regions as they are according to our 
present knowledge. The line separating confinement and Coulomb phases obtained 
in \cite{krw97} with $C_{\mbox{\scriptsize max}}$ has been supplemented by 
a point at $\lambda=1.35$ determined here and by the point $(1.4,0.52)$ found 
there only using $P_{\mbox{\scriptsize net}}$. It is seen that this transition 
line hits the boundary of the third phase at approximately $(1.43,-0.71)$. 

While $P_{\mbox{\scriptsize net}}$ has provided an unambiguous criterion
in the confinement phase and in the Coulomb phase, this is no longer so in 
the third phase. At fixed $(\lambda,\beta)$ in this region values 0 as well 
as 1 occur for $P_{\mbox{\scriptsize net}}$. The observations at
$(10,-1000)$ in \cite{krw97} appear to be related to this. That properties 
of monopole structures are no longer characteristic at sufficiently negative 
$\beta$ can be understood by noting that the respective quantities are
not invariant under the transformation $\beta \rightarrow -\beta$, 
$U_{\Box} \rightarrow -U_{\Box}$. 

As a characteristic feature of the third phase we have found that a number
of different states exists there between which transitions in the simulations 
are strongly suppressed. We have observed this phenomenon at various negative 
$\beta$ in the $\lambda$ range from 0 to 2.5 . Typical examples of time 
histories of the average plaquette 
$\epsilon=(1/6L^4)\sum_{\mu>\nu,x} (1-\cos \Theta_{\mu\nu,x})$
are given in Figure 2, with the results of seven simulation runs, exhibiting
different states and some transitions from lower to higher ones. It turns out 
that there is no correlation between the state reached and the type of start 
(hot or cold) of a simulation run. 

Comparing time histories of $\epsilon$ at different $(\lambda,\beta)$ the 
splittings of the states appear very similar. Closer inspection 
shows that at least nine states occur. The number of 
transitions between states increases with $\lambda$. They have so far always
been observed occurring from lower to higher plaquette values. The widths in
the time histories show little dependence on $\lambda$ while they strongly 
decreases if $\beta$ gets more negative (width$\times\beta$ being roughly 
constant). Thus at smaller negative $\beta$ it gets increasingly difficult to 
resolve splittings. 

The average $\epsilon$ found show not much dependence on $\lambda$, 
however, they increase as $\beta$ gets more negative. The $\epsilon$ 
values, being somewhat below 2, indicate that at sufficiently negative 
$\beta$ the average of $\cos \Theta_{\mu\nu,x}$ gets close to $-1$. 
That negative $\cos \Theta_{\mu\nu,x}$ occur at negative $\beta$ and 
positive $\cos \Theta_{\mu\nu,x}$ at positive $\beta$ in view of the
symmetry $\beta \rightarrow -\beta$, $U_{\Box} \rightarrow -U_{\Box}$ 
of the Wilson action is conceivable.

Though at negative $\beta$ the monopole density 
$\rho= (1/4L^4)\sum_{\rho,x} |M_{\rho,x}|$ appears of less interest, it
should be mentioned that the states are also seen in its time histories.
The respective time histories are correlated
with large values of $\rho$ corresponding to low values of $\epsilon$. 
The widths for $\rho$ decrease with $\lambda$. They are rather large so that, 
except at the largest values of $\lambda$, the resolution of the states 
is very poor. The value of $\rho$ decreases with $\lambda$.

The origin of the observed different states is not yet clear. Similar 
observations are made in simulations of spin glasses and of frustrated 
systems, and also in finite-temperature SU(N) gauge theory with states 
related to spontaneous $Z(N)$ breaking. In any case the third phase 
appears to have more complicated properties which deserve further 
(though computationally demanding) studies elsewhere.

\section{CRITICAL PROPERTIES} \setcounter{equation}{0}

\hspace{3mm}
Monte Carlo simulations with about $10^6$ sweeps have been performed 
at each of up to 20 values of $\beta$ in the critical region of the 
confinement-Coulomb phase transition at $\lambda=1.1$ and at $\lambda=0.8$
and for each lattize size considered.  Multihistogram techniques \cite{fs89} 
have been applied in the evaluation of the data. The errors have been
estimated by Jackknife methods.
In the finite-size analyses in addition to the specific heat and 
the Challa-Landau-Binder (CLB) cumulant \cite{binder} complex zeros of 
the partition function \cite{yl52}, in particular the Fisher zero $z_0$ 
closest to the $\beta$ axis \cite{f68}, have been used.

The specific heat is 
\be
C=\frac{1}{6L^4}\langle (E - \langle E \rangle)^2\rangle
\ee
where $E=\sum_{\mu>\nu,x} (1-\cos \Theta_{\mu\nu,x})$. 
For $d=4$ its maximum is expected to behave as 
\be
C_{\mbox{\scriptsize max}} \sim L^4
\ee
if the phase transition is of first order and as
\be
C_{\mbox{\scriptsize max}} \sim L^{\frac{\alpha}{\nu}}
\label{cm}
\ee
if it is of second order, where $\alpha$ is the critical exponent of the
specific heat and $\nu$ the critical exponent of the correlation length.

In Figure 3 we present the results for $C_{\mbox{\scriptsize max}}$ obtained 
on lattices with $L$ = 6, 8, 10, 12, 16 for $\lambda$ = 0.8, 0.9, 1.1. 
They include data from simulations of the present investigation 
($\lambda = 1.1$ and $\lambda = 0.8$) and ones from simulations 
of Ref.~\cite{krw97} ($\lambda = 0.9$). 
From Figure 3 it is already obvious that $\frac{\alpha}{\nu}$ as it occurs 
in (\ref{cm}) decreases with $\lambda$, which by the hyperscaling relation 
$\alpha = 2 - d\,\nu$ means that $\nu$ increases with $\lambda$. 

The fits to the data presented in Figure 3 give the values for 
$\frac{\alpha}{\nu}$ shown in Table I. They are clearly very far from 4
and thus the transition cannot be expected to be of first order. 
Using the hyperscaling relation $\alpha = 2 - d\,\nu$ the values for $\nu$ 
listed in Table II are obtained. They are different from the value 
$\frac{1}{2}$ of the gaussian case, however, rather close to it. Thus in
any case to conclude on second order appears quite safe. 

Similar results are obtained for the minimum of the CLB cumulant 
\ba
U_{\mbox{\scriptsize CLB}} &=& \frac{1}{3} (1-\frac{\langle E^4 \rangle}
  { \langle E^2 \rangle^2})
\label{fss5}
\ea
and for the imaginary part of the closest Fisher zero $z_0$. For these 
quantities finite-size scaling predicts the behaviors
\ba
  \mbox{Im}(z_0) &\sim& L^{-\frac{1}{\nu}} \quad ,\\
  U_{\mbox{\scriptsize CLB,min}} &\sim& L^{\frac{\alpha}{\nu}-4}
  \label{fss6} \quad .
\ea
The results of the respective fits are also listed in Tables I and II.
It is seen that the values obtained from different quantities roughly
agree, with some systematic deviations beyond the given statistical errors.
In any case it is obvious that $\nu$ increases somewhat with $\lambda$ and
that it is not far from $\frac{1}{2}$. 

The related critical $\beta_{\mbox{\scriptsize cr}}(L)$ (i.e.~the extrema
positions of $C$ and $U_{\mbox{\scriptsize CLB}}$ and the real part of
$z_0$) are given in Table III . They illustrate the dependence on
$\lambda$ and lattice size. The critical $\beta_{\mbox{\scriptsize cr}}(L)$
are expected to behave as
\be
\beta_{\mbox{\scriptsize cr}}(L)=\beta_{\mbox{\scriptsize cr}}^{\infty} 
+a L^{-\frac{1}{\nu}} \quad .
\label{beta}
\ee
Using the values of $\nu$ in Table II and the data for
$\beta_{\mbox{\scriptsize cr}}(L)$ in Table III, the numbers 
$\beta_{\mbox{\scriptsize cr}}(\infty)$ and $a$ in (\ref{beta}) have 
been calculated for $\lambda$=1.1. In Table IV they are compared with 
those obtained for $\lambda$=0.9 in \cite{krw97}. The errors given
are again only statistical ones.

The observed increase of $\nu$ with $\lambda$ could indicate a 
nonuniversal behavior with a maximal value (which could be even $\frac{1}{2}$)
reached at the boundary to the third phase. Another possiblity is that, 
because the observations are on finite lattices and the range of $\lambda$ 
considered is not too far from the first-order region, the starting point 
of the second-order region is not yet sharp. Then this effect should disappear 
on much larger lattices. 

To get hints on the dependence on lattice size effective $\nu$ (i.e.~a
sequence of ones related to two neighboring lattice sizes) have also been 
considered. Though there was a slight tendency in support of the above view
of a finite-size effect, it turns out that larger lattices and higher 
statistics (which unfortunately are computationally very expensive) would 
be needed for firm conclusions.

If for very large lattices $\nu$ gets independent of $\lambda$ for the 
$\lambda$ above a certain finite value and below the start of the third phase, 
the question arises what the ultimate $\nu$ would be. Because the 
numbers for $\nu$ in Table II come rather close to $\frac{1}{2}$ this could
be simply the gaussian value. If first order is confirmed along the boundary 
to the third phase, an interesting apect is that this transition line with 
$\nu=\frac{1}{4}$ is met by the second-order line with $\nu$ close to or 
equal to $\frac{1}{2}$.

\section*{ACKNOWLEDGMENT}

\hspace{3mm}
G.D. thanks A.~Weber for helpful discussions. W.K. is grateful to 
M.~M\"uller-Preussker and his group for their kind hospitality. 
This research was supported in part under DFG grant Ke 250/13-1.

\newpage

\renewcommand{\baselinestretch}{1.8} \small\normalsize

\begin{center}

{\bf TABLE I}

\vspace*{3mm}
Critical exponents
$\frac{\alpha}{\nu}$ from $C_V$, $U_{\mbox{\scriptsize CLB}}$.
\vspace*{3mm}

\begin{tabular}{|l|l|l|}
\hline
 $\lambda$ & $C_V$ &   $U_{\mbox{\scriptsize CLB}}$ \\
 \hline
  0.8      & 0.616(22)  &  0.756(23) \\
  0.9      & 0.485(35)  &           \\
  1.1      & 0.284(12)  &  0.391(12) \\
\hline
\end{tabular}
\vspace*{35mm}

{\bf TABLE II}

\vspace*{3mm}
Critical exponents
$\nu$ from Im($z_0$), $C_V$, $U_{\mbox{\scriptsize CLB}}$.
\vspace*{3mm}

\begin{tabular}{|l|l|l|l|l|}
\hline
 $\lambda$ &  Im($z_0$)  & $C_V$  & $U_{\mbox{\scriptsize CLB}}$ \\
 \hline
   0.8   &  0.404(5)    &   0.433(2)   & 0.421(3) \\
   0.9   &              &   0.446(5)   &          \\
   1.1   &  0.421(8)    &   0.467(2)   & 0.455(2) \\
\hline
\end{tabular}

\newpage

{\bf TABLE III}

\vspace*{3mm}
Critical $\beta_{\mbox{\scriptsize cr}}$ from $z_0$, 
$C_V$, $U_{\mbox{\scriptsize CLB}}$.
\vspace*{3mm}

\begin{tabular}{|l|r|l|l|l|}
\hline
$\lambda$ &  $L$ & Re($z_0$) & $C_V$  &  
$U_{\mbox{\scriptsize BCL}}$ \\
\hline
0.8 & 6 & 0.4637 (3)  & 0.4630 (3)  &  0.4647 (2) \\
    & 8 & 0.4783 (2)  & 0.4779 (2)  &  0.4783 (4) \\
    &10 & 0.4844 (2)  & 0.4841 (2)  &  0.4843 (2) \\
\hline
1.1 & 6   & 0.1409 (10)  & 0.1387(5)  & 0.1431 (6)  \\ 
    & 8   & 0.1636 (10)  & 0.1618(5)  & 0.1630 (6)  \\ 
    &10   & 0.1742 (4)   & 0.1731(5)  & 0.1736 (6)  \\ 
    &12   & 0.1775 (8)   & 0.1782(8)  & 0.1784 (8)  \\ 
    &16   & 0.1839 (4)   & 0.1837(4)  & 0.1838 (4)  \\ 
\hline
\end{tabular}

\vspace*{25mm}

{\bf TABLE IV}

\vspace*{3mm}
Fit data $\beta_{\mbox{\scriptsize cr}}^{\infty}$ and $a$ for $z_0$,
$C_V$, $U_{\mbox{\scriptsize CLB}}$.
\vspace*{3mm}

\begin{tabular}{|l|l|l|l|l|}
\hline
$\lambda$ &  & Re($z_0$)  & $C_V$       &   
$U_{\mbox{\scriptsize CLB}}$ \\
\hline
0.9 & $\beta_{\mbox{\scriptsize cr}}^\infty$ 
		    &     & 0.4059(5)    &    \\
    & $a$                 &     & -1.99(6)     &    \\
\hline
1.1 & $\beta_{\mbox{\scriptsize cr}}^\infty$ 
			  & 0.1888 (6)  &0.1902 (5)   &  0.1896 (5)     \\
    & $a$                 & -3.14(13)   & -2.41 (7)    & -2.49 (8)       \\
\hline
\end{tabular}

\end{center}

\newpage

\renewcommand{\baselinestretch}{1.5} \small\normalsize

\section*{Figure captions}

\begin{tabular}{rl}
Fig.~1. & Location of phase transition points $\beta_{\mbox{\scriptsize cr}}$
on $8^4$ lattice as function of $\lambda$\\& between confinement and Coulomb 
phases (circles from $C_{\mbox{\scriptsize max}}$, square from 
$P_{\mbox{\scriptsize net}}$)\\&and to third phase (diamonds). Curves drawn
to guide the eye.\\

Fig.~2. & Typical time histories of the average plaquette $\epsilon$
at negative $\beta$\\& obtained for seven different simulation runs,
\\& shown for $(\lambda,\beta)=(2,-20)$ and $8^4$ lattice. 
\\

Fig.~3. & Maximum of specific heat $C_{\mbox{\scriptsize max}}$ as function
of lattice size $L$\\&for $\lambda=0.8$, 0.9, and 1.1 at transition
points $\beta_{\mbox{\scriptsize cr}}$ between\\&
confinement and Coulomb phases.\\
\end{tabular}

\newpage

\centering

\begin{figure}[hb]
\epsfig{file=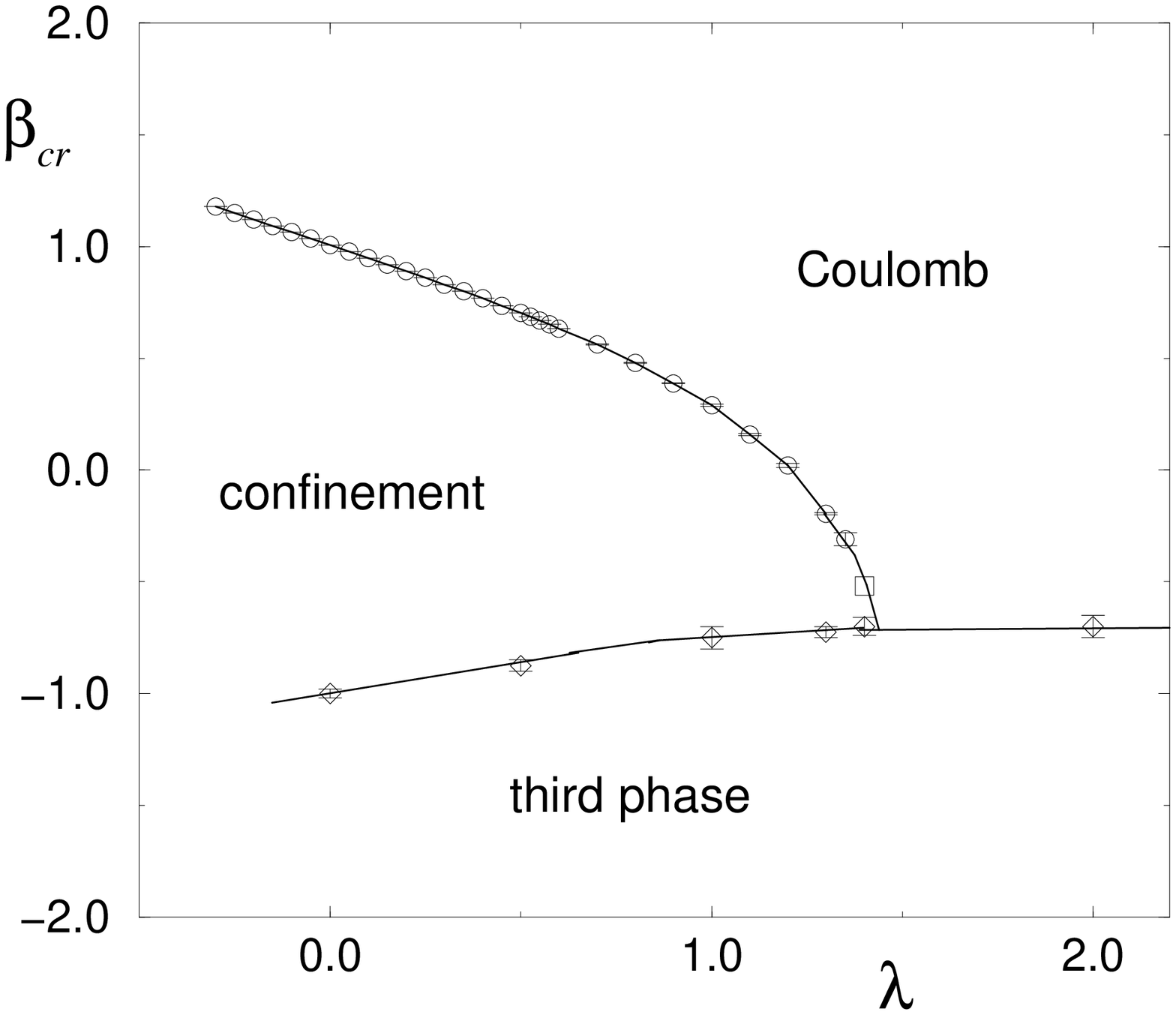,width=14cm,height=15cm}
\vspace*{10mm}
\caption{}
\end{figure}
  
\begin{figure}[hb]
\epsfig{file=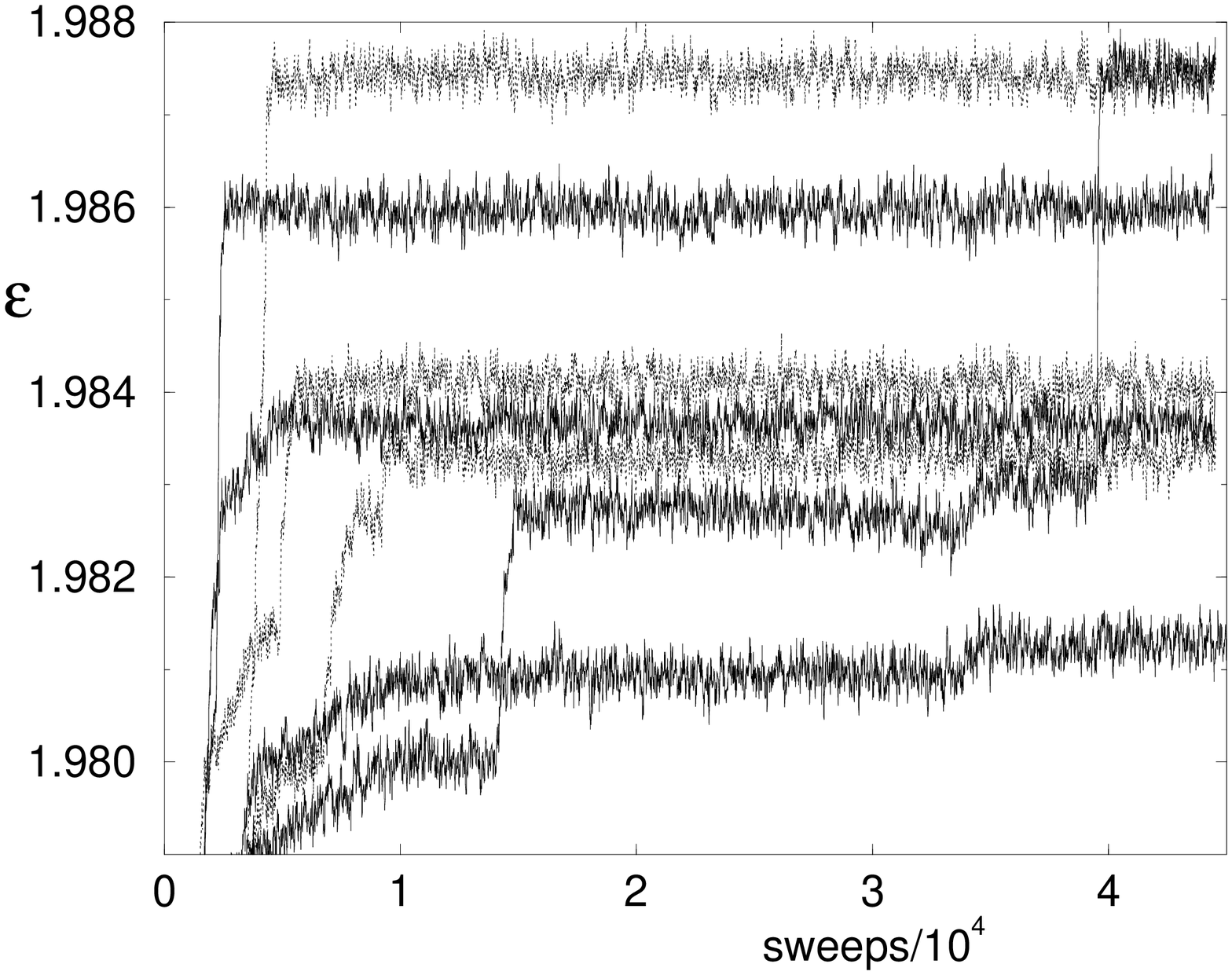,width=14cm,height=19cm}
\caption{}
\end{figure}
  
\begin{figure}[hb]
\epsfig{file=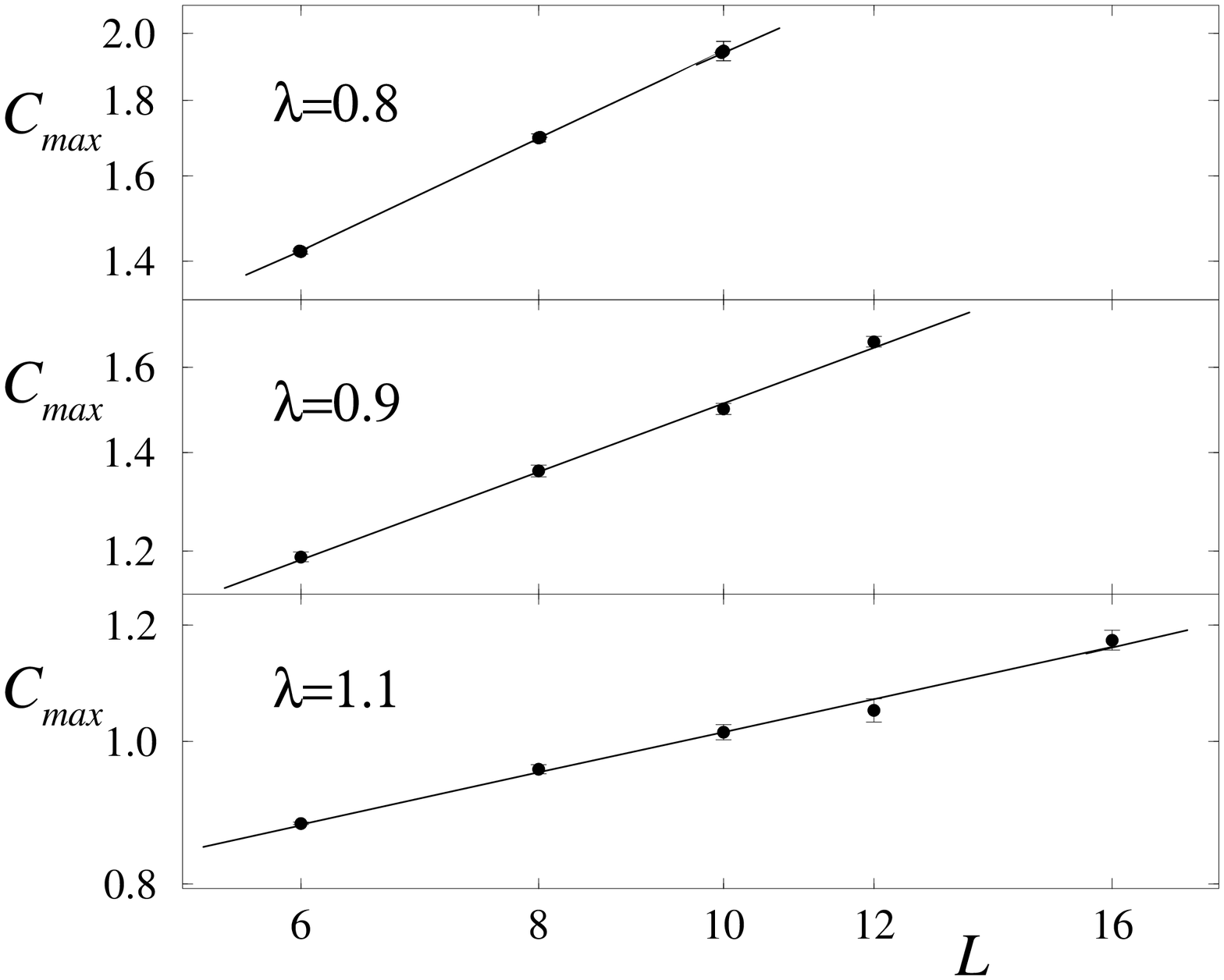,width=14cm,height=15cm}
\vspace*{10mm}
\caption{}
\end{figure}
  

\newpage

\begin{thebibliography}{9}

\bibitem{r97}C. Roiesnel, PC-561-09-97, hep-lat/9709081;
B. Klaus and C. Roiesnel, FUB-HEP/97-13, CPHT-S591.1297, hep-lat/9801036.
\bibitem{ejnz85}H.G. Evertz, J. Jers\'ak, T. Neuhaus, and P.M. Zerwas, Nucl.
Phys. {\bf B251}, 279 (1985).
\bibitem{jln96}J. Jers\'ak, C.B. Lang, and T. Neuhaus, Phys. Rev. Lett. 
{\bf 77}, 1933 (1996); Phys. Rev. D {\bf 54}, 6909 (1996).
\bibitem{cct97}I. Campos, A. Cruz, and A. Taranc\'on, DFTUZ 97/22, 
hep-lat/9711045; DFTUZ 98/08, hep-lat/9803007.
\bibitem{krw96}W. Kerler, C. Rebbi, and A. Weber, Phys. Lett. {\bf B380}, 346 
(1996).

\bibitem{krw94}W. Kerler, C. Rebbi, and A. Weber, Phys. Rev. D {\bf 50}, 6984 
(1994).
\bibitem{krw95a}W. Kerler, C. Rebbi, and A. Weber, Nucl. Phys. {\bf B450}, 452 
(1995).

\bibitem{bs85}J.S. Barber and R.E. Shrock, Nucl. Phys. {\bf B257}, 515 (1985).
\bibitem{dt80}T. DeGrand and D. Toussaint, Phys. Rev. D {\bf 22}, 2478 (1980).
\bibitem{krw97}W. Kerler, C. Rebbi, and A. Weber, Phys. Lett. {\bf B392}, 438 
(1997).
\bibitem{krw95}W. Kerler, C. Rebbi, and A. Weber, Phys. Lett. {\bf B348}, 565 
(1995).

\bibitem{hmp94}A. Hoferichter, V.K. Mitrjushkin, and M. M\"uller-Preussker,
Phys. Lett. {\bf B338}, 325 (1994).
\bibitem{bhmp96}Balasubramanian Krishnan, U.M.~Heller, V.K.~Mitrjushkin, and
M.~M\"uller-Preussker, HUB-EP-96/16, hep-lat/9605043.

\bibitem{fs89}A.M. Ferrenberg and R.H. Swendsen, Phys. Rev. Lett. {\bf 63},
 1195 (1989). 
\bibitem{binder}
 M.S. Challa, D.P. Landau, and K. Binder, Phys. Rev. B {\bf 34}, 1841 (1986).
\bibitem{yl52} C.N. Yang and T.D. Lee, Phys. Rev. {\bf 87}, 404 (1952).
\bibitem{f68}M.E. Fisher, in {\it Lectures in Theoretical Physics}, edited
by W.E. Brittin (Gordon and Breach, New York, 1964), Vol. VII C, p. 1.

\end{thebibliography}
\end{document}